\documentclass[aps,prd,twocolumn,superscriptaddress,bibnotes,showpacs,showkeys,floatfix]{revtex4-2}
\usepackage{graphicx}
\usepackage{amsmath}
\usepackage{lineno}
\usepackage{array}
\begin{document}
\sloppy
\title{Electron-antineutrino spectrum from precision measurements of $\rm{^{144}Pr}$-decay}

\author{\firstname{A.\,V.}~\surname{Derbin}}
\author{\firstname{I.\,S.}~\surname{Drachnev}}
\author{\firstname{D.\,V.}~\surname{Ivanov}}
\author{\firstname{I.\,M.}~\surname{Kotina}}
\author{\firstname{V.\,N.}~\surname{Muratova}}
\author{\firstname{N.\,V.}~\surname{Niyazova}}
\author{\firstname{D.\,A.}~\surname{Semenov}}
\author{\firstname{M.\,V.}~\surname{Trushin}}
\author{\firstname{E.\,V.}~\surname{Unzhakov}}
\affiliation{St.\,Petersburg Nuclear Physics Institute NRC Kurchatov Institute, 188350 Gatchina, Russia}

\date{\today}

\begin{abstract}
The $^{144}$Ce -- $^{144}$Pr electron antineutrino source is one of the most suitable radiochemical sources for experiments searching for the  oscillations of active neutrinos to the light sterile state.
In the current work the $\beta$-spectra of $^{144}\rm{Ce}$ -- ${^{144}\rm{Pr}}$ source  have been measured by two types of $\beta$-spectrometers based on silicon Si(Li) detectors in order to determine the energy spectrum of electron anti-neutrino emitted in the $\beta$-decay of $\rm{^{144}Pr}$ nuclei.
The nuclear shape factor of the ground state beta transition in ${^{144}\rm{Pr}}-{^{144}\rm{Nd}}$ has been obtained with high precision:  
$C(W) = 1 + (-0.0301 \pm 0.0007)W + (-0.101 \pm 0.006)W^{-1}$.
The reduced cross section for the inverse beta decay reaction on hydrogen for the $\rm{{^{144}Pr}}$ electron anti-neutrino source has been defined as $\rm{(4.7344 \pm 0.0006_{stat} \pm 0.013_{syst}) \times 10^{-44}~cm^2 {decay}^{-1}}$ that provides sufficient sensitivity to search for a sterile neutrino with a mass of $m_4 \sim 1$~eV and a mixing angle $Sin^2(2\theta_{14}) \sim 0.005$ using standard disappearance method.
\end{abstract}

\maketitle

\section{INTRODUCTION}
The three-flavor neutrino ($\nu_e, \nu_\mu, \nu_\tau$) oscillations have been reliably observed using different neutrino sources and detectors.
This fundamental discovery has demonstrated the presence of nonzero neutrino masses and has already went beyond the Standard Model (SM).
Three mixing angles ($\theta_{12}, \theta_{23}, \theta_{13}$) and two mass difference squares  $\Delta m_{12}^2, \Delta m_{23}^2$ of the Pontecorvo-Maki-Nakagawa-Sakata (PMNS) matrix have already been measured with acceptable accuracy~\cite{Nav2024}.
However, the possible existence of additional sterile neutrino that mixes with the three SM neutrinos has remained one of the crucial unresolved problems of particle physics since the beginning of the century.
The results of several neutrino experiments can be interpreted as indications of the light sterile neutrinos existence with $\Delta m_{14}^2$ of around $1~\rm{eV^2}$~\cite{Ace2023}.
\par
The LSND experiment was the first one to observe $\approx 3\sigma$ excess of electron anti-neutrinos in the beam of neutrinos produced by pions decay at rest~\cite{Ath1996, Agu2001}.
The LSND results were tested by the MiniBooNE experiment which has also reported a low energy excess in $\nu_\mu \rightarrow \nu_e$ and $\bar\nu_\mu \rightarrow \bar\nu_e$ channels in neutrino beam from decay-in-flight pions.
The overall significance of the MiniBooNE excess in both neutrino and antineutrino modes is $4.8~\sigma$~\cite{Agu2009,Agu2018,Agu2021}.
The positive results of accelerator-based neutrino experiments LSND and MiniBooNE are currently being tested by JSNS$^2$~\cite{Lee2023} and MicroBooNE~\cite{Abr2022,Abr2022a} collaborations. 
\par
The ``reactor antineutrino anomaly'' (RAA) \cite{Men2011} has shown a $(5-6)\%$ discrepancy between the expected (based on a new anti-neutrino flux calculations~\cite{Mue2011,Hub2011}) and the observed count rates of inverse beta-decay (IBD) reaction at the detectors installed close, $10-30$~m, to the nuclear reactors. 
This effect has been explained via mixing between electron and sterile (anti)neutrino with oscillation parameters  $\Delta m_{14}^2 \approx 1 \rm{eV^2}$ and $sin^2(2\theta_{14}) \approx 0.05$~\cite{Den2017, Gar2018}. 


However, the recent reactor experiments~\cite{Ade2019,Bak2019,Alm2023} and new models of reactor neutrino spectra~\cite{Kop2021}, based in particular on the so-called summation approach, made the discrepancy between measurements and theoretical predictions less significant~\cite{Giu2021,Per2023}. 


\par
The next indication of the possible existence of a sterile neutrino has been related with $\approx 16\%$ ($2.9\sigma$) lack of count rate in Ga-Ge radiochemical solar neutrino detectors SAGE~\cite{Abd1999,Abd2006} and GALLEX/GNO~\cite{Ham1998,Alt2005} during calibration with intense artificial neutrino sources $^{51}$Cr and $^{37}$Ar.
Recently, BEST collaboration has presented the new results obtained with $\rm{{^{51}Cr}}$ source and Ga-Ge detector and the deficit of the observed $\rm{^{71}Ga}(\nu_e,e^-)\rm{^{71}Ge}$ reaction events stands at more than $5\sigma$ level~\cite{Bar2022, Bar2022a}.
\par
Several experiments with electron antineutrino sources  ($\rm{^{144}Ce}$, $\rm{^{106}Ru}$, $\rm{^{90}Sr}$ and  $\rm{^{42}Ar}$) were proposed to verify the detected anomalies by looking for neutrino oscillations at  short distances~\cite{Gri2011, Bel2013, Gan2014}. 
One of these experiments was Borexino-SOX (Short distance neutrino Oscillations with BoreXino) in the Gran Sasso underground laboratory (Italy) with electron antineutrino ($\bar{{\nu}}_e$) $\rm{{^{144}Ce} - {^{144}Pr}}$ source~\cite{Bel2013, Not2015, Alt2016, Pal2017, Gro2017}. 
The unprecedentedly low radioactive background of the Borexino detector, good energy and spacial resolutions together with large size allowing to perform an oscillometry (wave) measurement within the detector volume, and the well demonstrated capability to detect low energy electron antineutrinos have represented an ideal case for the study of short distance neutrino oscillations with artificial sources.
Unfortunately, it was decided to suspend the Borexino\_SOX experiment, 
however, the source $\rm{{^{144}Ce} - {^{144}Pr}}$ remains one of the most promising for the future experiments searching for light sterile neutrinos.

The $^{144}\rm{Ce}$ nucleus undergoes three first-order forbidden non-unique beta-decays to the ground ($Q_\beta=318.7\,(8)$~keV, $76.5$\%) and excited levels $1^-$ ($Q_\beta=238.6$~keV, $3.9$\% and $Q_\beta=185.2$~keV, $19.6$\%) of the $^{144}\rm{Pr}$ nucleus.

Decays of the $^{144}\rm{Pr}$ to the ground ($Q_\beta=2997.5\,(24)$~keV, $97.9$\%) and first excited ($Q_\beta=2301.0$~keV, $1.04$\%) states of the $^{144}\rm{Nd}$ nucleus correspond to first-order forbidden non-unique and unique transitions, respectively. 
Decay into the $1^-$ level of $\rm{{^{144}Nd}}$ with an energy of $2185$~keV ($Q_\beta=811.8$~keV, $1.05$\%) is an allowed transition, the fitting of the beta spectrum does not require additional shape corrections as in the case of forbidden transitions.

The main objective of $\rm{^{144}Pr}$-decay study is to determine the spectrum of electron antineutrinos in the decay of $\rm{{^{144}Pr}}$ with an energy above $1806$~keV, the threshold of the inverse beta decay reaction on hydrogen.
This is possible at electron energies less than $1192$~keV in the case of a transition to the ground state and at an electron energy less than $495$~keV in the decay to the first excited state.

The expected sensitivity of the experiment with $\rm{{^{144}Ce} - {^{144}Pr}}$ source relative to the sterile neutrino oscillation parameters depends on the accuracy of flux determination as well as the energy spectrum shape for the electron anti-neutrino emitted in the $\beta-$decay of $\rm{{^{144}Pr}}$ nuclei.
The sensitivity 
is evaluated by comparing the observed event rate and positron spectrum of IBD reaction, binned as a function of both energy and distance, with respect to the expected distribution in the presence of oscillations. 
Three data analyses can be performed: these are the so-called ``count rate'', ``shape'' and ``rate + shape'' tests. 
To illustrate the importance of the spectral shape of the antineutrino spectra measurement, just note that the existing shape spectrum  uncertainties translates into a $\sim 10$\% uncertainty on the predicted IBD rate~\cite{Gan2013, Gaf2015, Viv2016, Alt2016}.

The purpose of this work was a precise measurement of the $\beta-$spectra from $\rm{{^{144}Ce} - {^{144}Pr}}$ source and determination of the spectrum of electron anti-neutrinos emitted in the decay $\rm{{^{144}Pr}}$ nuclei with great precision.

\section{EXPERIMENTAL SETUP}
The spectra of the $\rm{{^{144}Ce} - {^{144}Pr}}$ source were measured by two setups: one using a beta-spectrometer in the classical "target - detector" configuration, and the other relies an originally developed total absorption $4\pi$-spectrometer consisting of a pair of Si(Li)-detectors. 
These measurements correspond to the application of two approaches to determining the detector response function - accurate Monte Carlo modeling and the use of a detector response function close to a Gaussian function.

\subsection{Spectrometer ``target-detector'' type}
\begin{figure}[t]
\includegraphics[bb = 100 80 500 755, width=7.5cm,height=10.5cm]{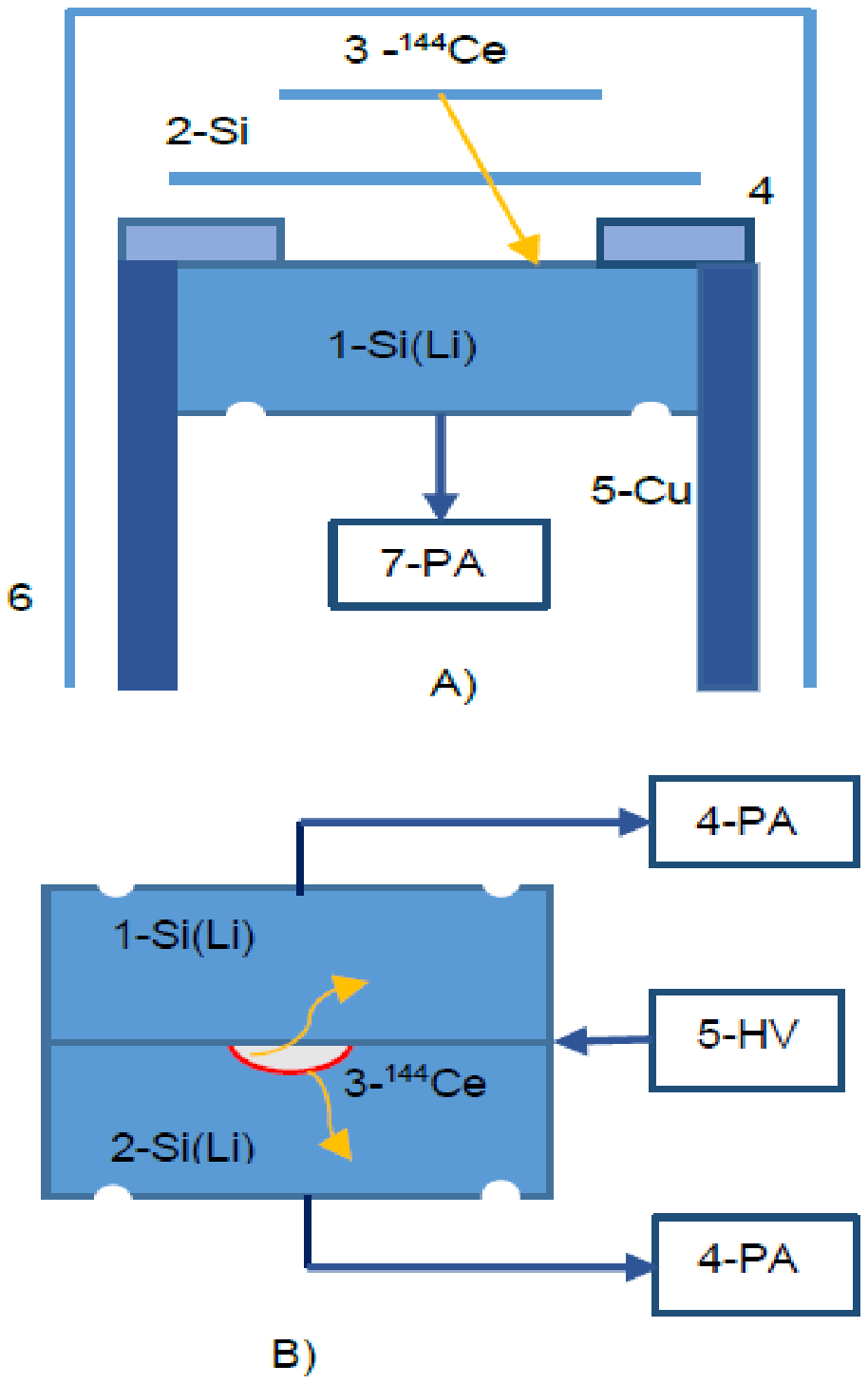}
\caption{Schematic representation of the spectrometers. 
A)``target - detector'' spectrometer: 1 - Si(Li) full absorption detector, 2 - Si transmission detector, 3 - $\rm{^{144}Ce-^{144}Pr}$ source, 4 - tungsten collimator, 5 - Cu cold finger, 6 - vacuum cryostat. B) $4\pi-$spectrometer: 1,2 - Si(Li) full absorption detectors, 3 - $\rm{^{144}Ce-^{144}Pr}$ source, 4 - preamplifiers, 5 - bias voltage.}
\label{Fig1}
\end{figure}

The ``target-detector'' type spectrometer was created on the basis of a semiconductor Si(Li) detector with a thickness of $10.2$~mm and the sensitive area diameter of $20$~mm, which had the ``top-hat'' geometry~(Fig.~\ref{Fig1}(A)).
These dimensions ensure an effective absorption of electrons with energies up to $3$~MeV.
\par
The Si(Li)-detector produced from p-type single-crystal silicon with a resistivity of $4$~k$\rm\Omega~\rm{cm}$ and a carrier lifetime of $800$~$\rm{\mu}$s using standard technology that has been developed and tested by the Petersburg Nuclear Physics Institute. 
The characteristics of such detectors were described in~\cite{Baz2018,Ale2018,Ale2020, Ale2023}.
\par
The Si(Li)-detector was placed in a holder and equipped with a tungsten collimator with diameter of $12$~mm.
The total insensitive layer of the input detector window including the gold-plated $p$-contact was about 470~nm of silicon equivalent.
The thickness of the insensitive layer of diffuse lithium on the back side of the detector was identified through the ratio of the total absorption peak areas for $^{241}$Am source gamma lines with energies of $26.2$~keV and $59.6$~keV to be $420$~$\mu\rm{m}$. 
\par
The removable square-shaped transmission detector $\rm{24\times24~mm^2}$ has a thickness of $300$~$\mu\rm{m}$.
The energy resolution of the cooled detector was measured to be FWHM = 7.9~keV.
Switching the transmission detector into anti-coincidence mode with the Si(Li)- detector allows one to determine the $\gamma$- and X-ray contribution to the Si(Li) detector spectrum \cite{Ale2018}.
\par
The $^{144}$Ce source in the form of a dried drop of colloidal solution was placed on the surface of a mylar substrate with thickness of $50$~$\mathrm{\mu}$m. 
The substrate was fixed in a delrin ring and placed on a supporting structure at the distance of $8.9$~mm from the detector surface. 
The thickness distribution over the spot of the dried source $^{144}$Ce was determined from the measured energy spectrum of alpha particles originating from the decays of $^{241}$Am and $^{244}$Cm, which were preserved in the source as low activity impurities after purification of the $^{144}$Ce source. 
The average thickness of the source was determined as~$\approx 1.0~\mu$m.
\par
A HPGe-detector 160 $\rm{cm^3}$ volume in a low background setup was used to detect additional $\gamma$-activity in the $\rm{^{144}Ce}$-source. 
The main extraneous $\gamma$'s are associated with the isotope  $\rm{^{154}Eu}$ activity, which was 0.004 relative to the $\rm{^{144}Ce}$ activity at the end of the measurements. 
The comparison was made using $\gamma$-lines with energies of 723~keV and 697~keV for $\rm{^{154}Eu}$ and $\rm{^{144}Pr}$, respectively. 
The next most important activity of $\rm{^{152}Eu}$ is an order of magnitude less than the activity of $\rm{^{154}Eu}$.
\par
The entire installation of the Si(Li)-detector and ${\rm{^{144}Ce}}$ source was placed in a vacuum cryostat and was cooled down to the temperature of liquid nitrogen (LN). 
The Si(Li)-detector was equipped with a charge-sensitive preamplifier with a field-effect transistor placed inside the vacuum cryostat and cooled down to the LN temperature.
The negative bias voltage of $1$~kV was applied directly to the gold coating of the detector.
\par
The signal from the preamplifier was registered by 8-channel CAEN v1725 digitizer with sampling rate of $250$~MHz.
Further transformations with the signal were performed digitally. 
The time reference and the record trigger of the event were formed on the basis of the signal after digital CR-2RC formation, while the amplitude analysis was performed using triangular shaping. 
The signals were recorded on event-by-event basis, a time reference and amplitude were recorded for each event. 
\par
The setup also included a scintillation detector based on a 3" ($2.5$~kg) bismuth orthogermanate (BGO) crystal coupled to a Hammamatsu R1307 PMT, the output of which was connected to the digitizer channel. 
Time and amplitude information was recorded for the signal of the BGO detector, which made it possible to analyze time-amplitude coincidences allowing to discriminate the $\beta-$spectrum of the allowed transition ${^{144}\rm{Pr}~(0^-)} \rightarrow {^{144}\rm{Nd}~(1^-)}$. 
The time resolution of the spectrometric paths was about $\sim100$~ns.

In order to select transitions to the $1^{-}$~state, we set the condition on the energy of $\gamma$-quantum registered by BGO detector to be above 1~MeV. 
This condition is fulfilled for the $\gamma$-quanta from the full absorption peak and part of the Compton scattering tail  and allows for exclusion of transitions to $2^+$~state followed by a single 697 keV $\gamma$-quantum. 
The energy resolution of the BGO detector allows separating reliably the energies of 697~keV and 1~MeV. 

The spectrum of Si(Li) detectors in coincidence with BGO signals that exceeded 1 MeV consists of the $\beta$-spectrum of the allowed transition of ${^{144}\rm{Pr}~(0^-)}$ to the excited state of ${^{144}\rm{Nd}~(1^-)}$ and random coincidences.
As will be shown below the spectrum shape is in good accordance with theoretical shape of the allowed $\beta$-transition with 812~keV endpoint.
The choice of relatively small BGO-detector was driven by requirement for high $\gamma$-registration efficiency in case of low background of random coincidences.
The agreement of the total measured spectrum shape to the shape of the allowed $\beta$-transition represents an important validity criterion for the measurements, response function and fitting procedure.

\begin{figure}[ht]
\includegraphics[bb = 30 100 600 755, width=8cm,height=10.5cm]{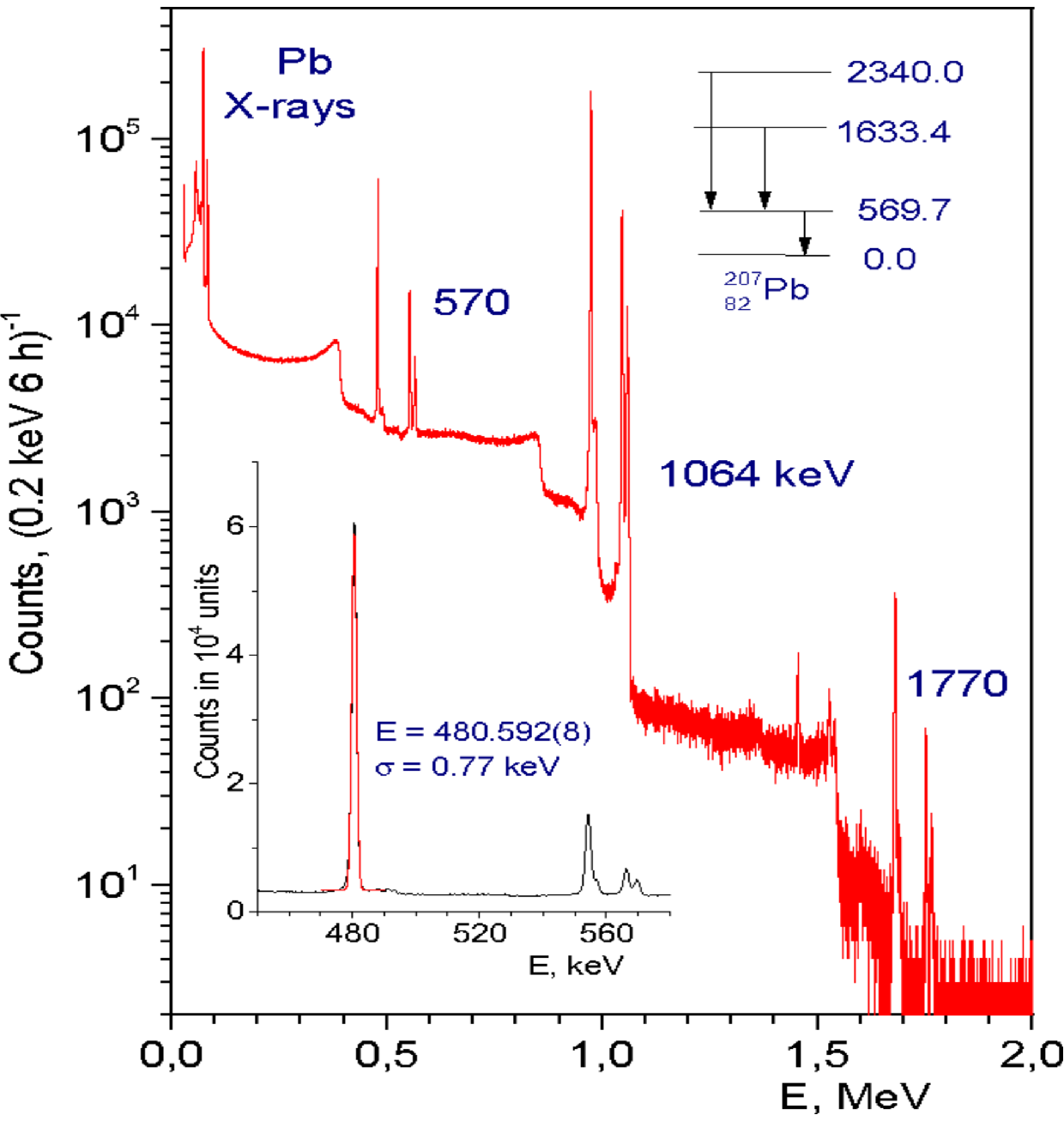}
\caption{Energy spectra in the ranges $(0.01-2.0)$~MeV measured by a Si(Li)-detector with a $\rm{^{207}Bi}$ source. The bottom insert shows the electron peaks corresponding to the internal conversion from K-, L- and M-shells while the $569.7$~keV nuclear level discharge. The nuclear levels of  $\rm{^{207}Pb}$ are shown in the top inset.}\label{Fig2}
\end{figure}

\begin{figure}[ht]
\includegraphics[bb = 30 80 600 755, width=9.5cm,height=10.5cm]{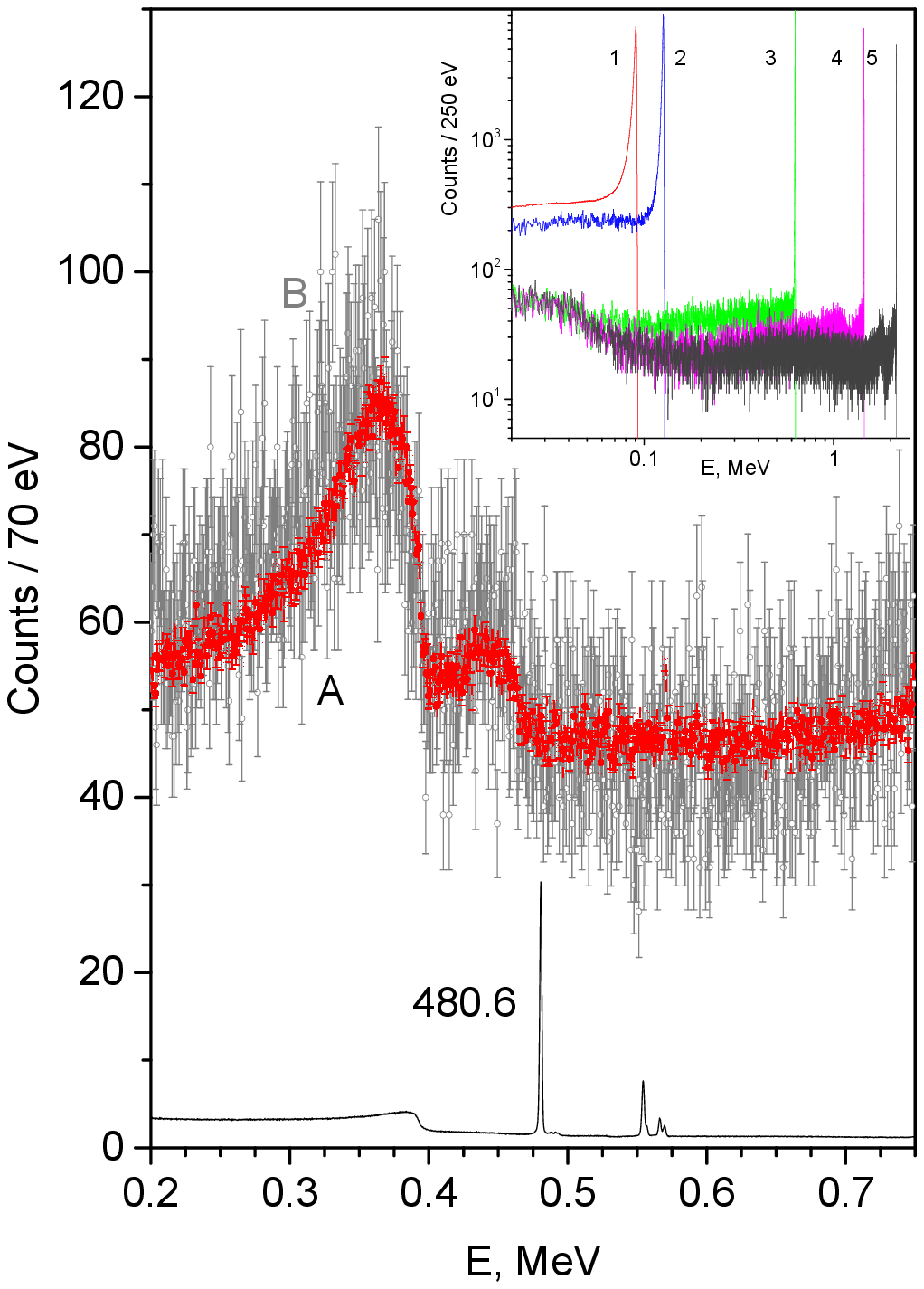}
\caption{The spectrum of $\rm{^{207}Bi}$ measured in coincidence with transition detector (black) and simulated with Monte Carlo method (red). The lower plot shows the full absorption detector spectrum obtained without transition detector. The inset shows detector response functions for $\rm{^{144}Ce – ^{144}Pr}$ conversion electrons with energies 91.5 keV (1), 127 keV (2), 626 keV (3), 1446 keV (4) and 2145 keV (5) \cite{Ale2018}.}
\label{Fig1a}
\end{figure}

The energy calibration was performed with help of total absorption peaks for gamma quanta with energies of $570$~keV and $1063$~keV of the $^{207}$Bi source~(Fig.~\ref{Fig2}, \cite{Baz2018}). 
The energy resolution FWHM (full width at half maximum), determined by the conversion electron peak with an energy of $480.6$~keV, was $1.8$~keV ($\sigma = 770$~eV).
The results of Monte Carlo simulation of the calibration spectrum show good agreement with the measured spectrum, including the electron backscattering, edges and shape of Compton scattering \cite{Ale2018}.

The spectrum of $\rm{^{207}Bi}$ measured by Si(Li)-detector in coincidence with transition detector and the spectrum simulated with Monte Carlo method are shown in Fig.~\ref{Fig1a} \cite{Ale2018}. 
When passing through a thin 300 $\mu$m detector, conversion electrons lose energy in accordance with the Landau distribution and form two broad peaks in the Si(Li)-spectrum, shifted by an energy of about 100 keV.
The lower plot shows the full absorption detector spectrum obtained without transition detector. 
The Si(Li)-detector response functions calculated by Monte Carlo method using the real "target - detector" geometry of  for electrons with different energies are also shown. 

Measurements were carried out continuously for 2024 hours, with short technological stops.
The long duration of the measurements was defined by the low activity of the $^{144}$Ce source, which was $\sim 15$~Bk.
The drift of the gain coefficient of the Si(Li) detector spectrometric channel during the long-term measurements was compensated by re-calibrations driven by the peak of conversion electrons with an energy of 91.5~keV at decays of $^{144}$Ce and the 5.902~MeV $\alpha$-peak of $^{244}$Cm, which were present in the measured spectra. 
\par
The ``target - detector'' type spectrometer has a number of undoubted advantages, in particular, it allows one to measure $\beta$-spectra  of various sources using the same detector with quick replacement of the sources \cite{Ale2020}.
The main problem with the scheme is the backscattering of electrons from the crystal surface, which depends on the angle of incidence and leads to a large low-energy "tail" in the detector response function.

\subsection{4$\pi$-spectrometer}
The original spectrometer with 4$\pi$ geometry was designed on the basis of two Si(Li) detectors with the thicknesses of about $9$~mm each, which exceeds the ionisation path of electrons with an energy of $3$~MeV. 
The detectors were specially manufactured for this experiment and they were made in the ``top-hat'' shape with external diameters of $27$~mm and $23$~mm, height of $10$~mm and diameters of the sensitive area of $20$~mm and $18$~mm.
The Si(Li) detectors had different external diameters for the convenience of assembling the spectrometer with two detectors docked closely~(Fig.~\ref{Fig1}, (B))~\cite{Ale2023, Ale2021,Ale2021a,Bak2022}.

The characteristics of the detectors were tested in a separate vacuum cryostat using $\gamma$-, X-rays, conversion and Auger electrons from the $\rm{{^{207}Bi}\rightarrow{^{207}Pb}}$ source.
The energy resolution  measured for conversion electrons with energy of 481~keV from  $^{207}$Bi source was $\rm{FWHM} \approx 2.0$~keV for both detectors.
\par
The total thickness of the insensitive layer of the Si(Li) detector, contributed by the deposited layers of palladium and gold and the surface layer of silicon, corresponds to a thickness of about 500~nm in silicon equivalent. 
Passing through such a thickness, electrons with energies of $20$~keV and $3$~MeV lose about $1$~keV and $0.1$~keV respectively.
\par
Comparison of the measured intensities of X-ray K$_{\alpha 1}$, K$_{\alpha 2}$ and K$_{\beta 123}$ lead peaks with the results of Monte Carlo calculations using GEANT4.10.6 package allowed us to determine the thickness of the detector sensitive i-region. 
The useful thickness of both detectors determined in this way exceeds $8.5$~mm, which ensures, if one does not takes into account the backscattering, the complete absorption of any electrons with the energies less than $3.3$~MeV.
\par
In the center of the front surface of one of the detectors a small cavity with a diameter of $5$~mm and a depth of $1$~mm was made by chemical etching. 
The $^{144}$Ce $\beta$-source  was placed into the cavity, directly onto the gold coating of the planar Si(Li) detector. 
The second detector was located on the detector with a cavity without any gap, and the bias voltage was applied to the resulting joined metal contact.
\par
The assembly was placed in a vacuum cryostat and cooled down to the temperature of liquid nitrogen. 
The spectrometric channels were completely similar to the ones used in the ``target-detector'' type spectrometer and included preamplifiers with cooled field-effect transistors, the outputs of the preamplifiers were connected to the CAEN v1725 digitizer.

The total registered energy spectrum is obtained by summing individual spectra from each of the Si(Li)-detectors and the energy spectrum recorded by both detectors in coincidence.
The obtained spectrum solves the response function problem associated with backscattering of electrons from the detector surface.

The BGO-detector placed on the cryostat cap at a distance of $~25$~mm from the $\rm{{^{144}Ce}-{^{144}Pr}}$ source was also connected to one of the inputs of the digitizer.
\par
The proposed full absorption original $4\pi$-spectrometer can be used for direct measurements of $\beta$-specta and does not require corrections of the response function for the electrons backscattering from the crystal surface.

\section{DATA ANALYSIS}



The energy spectra of particles arising from the $\beta-$decay of $\rm{{^{144}Ce}}$, which are $\beta$-, Auger-, and conversion electrons, X- and $\gamma$-rays, have a complex form for fitting.
Therefore, we will precisely fit the $\beta$-spectrum of $\rm{{^{144}Pr}}$ in the range of ($250 - 3100$)~keV, capturing a small region of the spectrum of $\rm{{^{144}Ce}}$ near the end-point energy $Q_{\beta}(\rm{{^{144}Ce}})$, in order to use it as an additional reference energy point. 
To determine the neutrino spectrum at neutrino energies greater than  ($Q_{\beta}(\rm{{^{144}Pr}})-250~\rm{keV}$), extrapolation of the obtained $\beta$-spectrum to the low-energy region will be used.

In our work the detailed expression for electron energy spectrum of the given $\beta$-transitions included the following components \cite{Wil1990,Beh1971,Beh1983,Hub2011,Hay2018}:
\begin{eqnarray} \label{eq:beta_basic}
    S(W)dW &=& K\,pW\,(W - W_0)^2\,F(Z,W) \cdot\\
        &\cdot&L_0(Z,W)\,S(Z,W)\,G_{\beta}(Z,W)\,C(Z,W)\,dW \nonumber
        \end{eqnarray}
where $K$ is a normalization constant, $p$~is electron momentum, $W$~is total electron energy in units of electron mass ($W = T/mc^2 + 1$, $T$ -- kinetic energy of electron), $W_0$~corresponds to the endpoint energy of $\beta$-spectrum, $Z$~is electric charge (i.e. proton number) of the daughter nucleus, and $F(Z,W)$ is Fermi function.

The first part $pW(W - W_0)^2$ of expression~(\ref{eq:beta_basic}) is phase space factor accounting for distribution of decay energy between produced particles (i.e. electron and antineutrino).
The second factor $F(Z,W)$ is so called Fermi function describing the Coulomb interaction of outgoing electron with daughter nucleus.
Fermi function originates from solution of Dirac equation in assumption of point-like and infinitely heavy nucleus and in fact represents the leading order QED correction for description of $\beta$-decay.

In case of high precision measurements the basic Fermi approximation becomes insufficient and one has to take more intricate effects into account, which is achieved by introducing additional electromagnetic and weak corrections accounting for finite size and mass of the daughter nucleus ($L_0(Z,W)$)~\cite{Wil1990}, screening of nuclear electric field by electrons of atomic shell ($S(Z,W)$)~\cite{Beh1983,Hub2011}, and radiative correction due to emission of photons by charged particles involved in $\beta$-decay ($G_\beta(Z,W)$)~\cite{Sir1967,Hay2018}.

The screening correction $S(Z,W)$ accounts for the contribution of atomic electrons still remaining in ``parent'' configuration affecting the Coulomb potential of the ``daughter'' nucleus produced in $\beta$-decay.
The initial work around for this problem suggested simple rescaling of the total energy of a $\beta$-particle $\bar{W} = W - V_0$ \cite{Ros1936}, where $V_0$ is an effective shift of Coulomb potential due to the screening of atomic electrons.
While this approach has been proven to be adequate for energies $\gg V_0$ \cite{Mat1966}, due to the energy rescaling this technique becomes invalid in low-energy part of the spectrum when $W \le V_0$, exactly where it is of the most importance for evaluation of the high-energy end of corresponding neutrino spectrum.

In case of Ce and Pr atoms the screening shift $V_0 \approx 0.01$~MeV~\cite{Beh1983}.
Since realistic description of screening correction below $V_0$ introduces significant complexity, we performed the upper estimate of potential uncertainty caused by screening effect by performing calculations for two border cases: no screening below $V_0$~($F_{scr}(W)=1$) and maximum screening below $V_0$ ($F_{scr}(W)=F_{scr}(V_0)$).
These two options introduce the greatest uncertainty into the neutrino spectra and, as a consequence, into the expected count rate of the inverse beta decay events and were considered as the largest systematic uncertainty.

The radiative correction to the $\beta$-spectrum at order $\alpha$ has been computed in \cite{Sir1967,Sir1978,Sir2011}.
The radiative corrections have two components connected with the virtual photons on the Feynman lines of the charged particles and the emission of real internal bremsstrahlung (IB) photons~\cite{Bat1995}.
 As a result, when reconstructing the antineutrino spectrum from the $\beta$-spectrum, the correction $G_{\beta}(Z,W)$ is replaced not by $G_{\beta}(Z,W_0-W)$, but a new correction $G_{\nu}(Z,W_\nu)$~\cite{Bat1995,Sir2011}.
The $\beta$-spectrum and the antineutrino spectrum are described by two different radiative corrections $G_\beta(Z,W_e)$ and $G_\nu(Z,W_\nu)$, respectively.
This approach should be used when determining the antineutrino spectrum based on "target-detector" spectrometer data since only $8.5$\% of the IB photons are emitted in the Si(Li)-detector solid angle.

The peculiarity of the described above $4\pi$-spectrometer is that it registers with high efficiency the IB photons with energy up to 100 keV, which are considered as part of the radiative corrections. 
For example, 1~MeV electron appearing in $\rm{^{144}Pr}$-decay emits $75$\% of the IB photons with an energy of less than $100$~keV, of which only $6$\% leave the $4\pi$-detector without interaction.
As a result, we used two different forms of the radiative correction function fitting the data of ``target-detector'' and $4\pi$-spectrometer. 
For the $4\pi$-spectrometer we used the correction $G_\nu(Z,W_\nu)$ introduced for the neutrino spectrum \cite{Sir2011,Bat1995} with the natural replacement of $W_\nu$ by $(W_0-W_\nu)$.

In addition to electromagnetic and kinematic effects, another higher order correction to Fermi function arise from internal nuclear structure and matrix elements of a given $\beta$-transition.
Nuclear shape-factor $C(Z,W)$ effectively accounts for complex collective nuclear interactions and becomes significant in case of forbidden transitions \cite{Hay2018}.
It is assumed that for allowed transitions $C(Z,W)=1$, but except for some trivial cases of allowed and unique-forbidden transitions, analytical evaluation of nuclear shape-factor appears to be very difficult, thus leaving the experimental extraction the only viable option.

As noted above, we need to extrapolate the shape-factor $C(Z,W)$ to the region of electron energies below $250$~keV. 
The encouraging fact is that the expected theoretical dependence of $C(Z,W)$ for the first-order forbidden non-unique $\rm{{^{144}Pr}(0^-)\rightarrow{^{144}Nd}(0^+)}$ transition  $C(Z,W)\sim (p^2_e + E^2_\nu+2\beta^2 E_\nu E_e)$ is almost constant in this region, smoothly increasing by only $~7\%$.

The weak magnetism correction is not included in equation (\ref{eq:beta_basic}) because it is vanished for the main decay branch of $\rm{^{144}Pr}$-nucleus~\cite{Hay2014}. 




\subsection{Analysis of the data of spectrometer ``target-detector'' type}

 \begin{figure}[t]
  	\includegraphics[width=9cm,height=10.5cm]{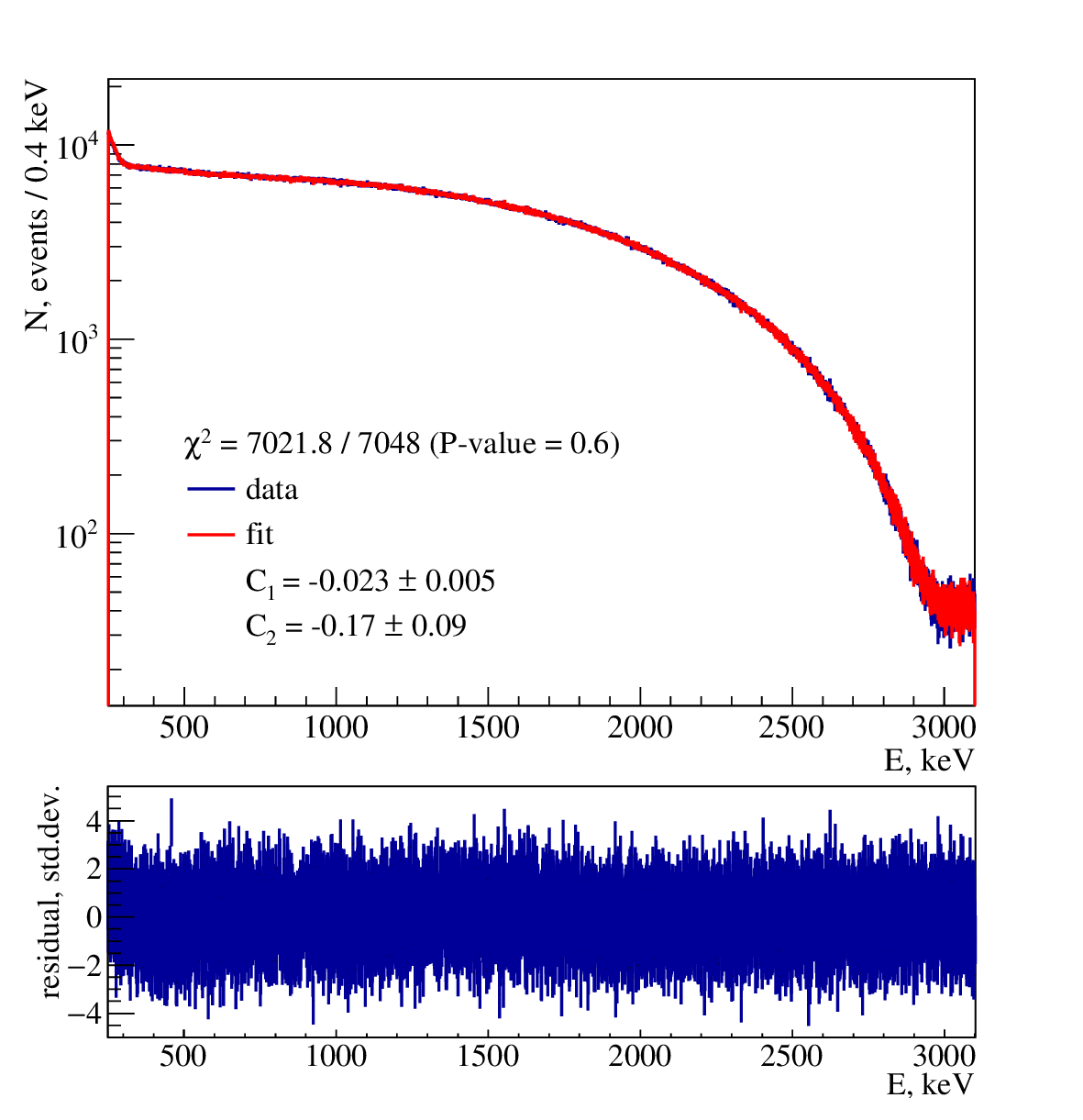}
  	\caption{The measured spectrum of the $^{144}$Ce-$^{144}$Pr source with the spectrometer in the ``target-detector'' geometry. The fitting result is shown by red line. The difference is given in the units of standard deviations (SD).}
  	\label{Fig3}
 \end{figure} 
\begin{figure}[t]
	\includegraphics[width=9cm,height=10.5cm]{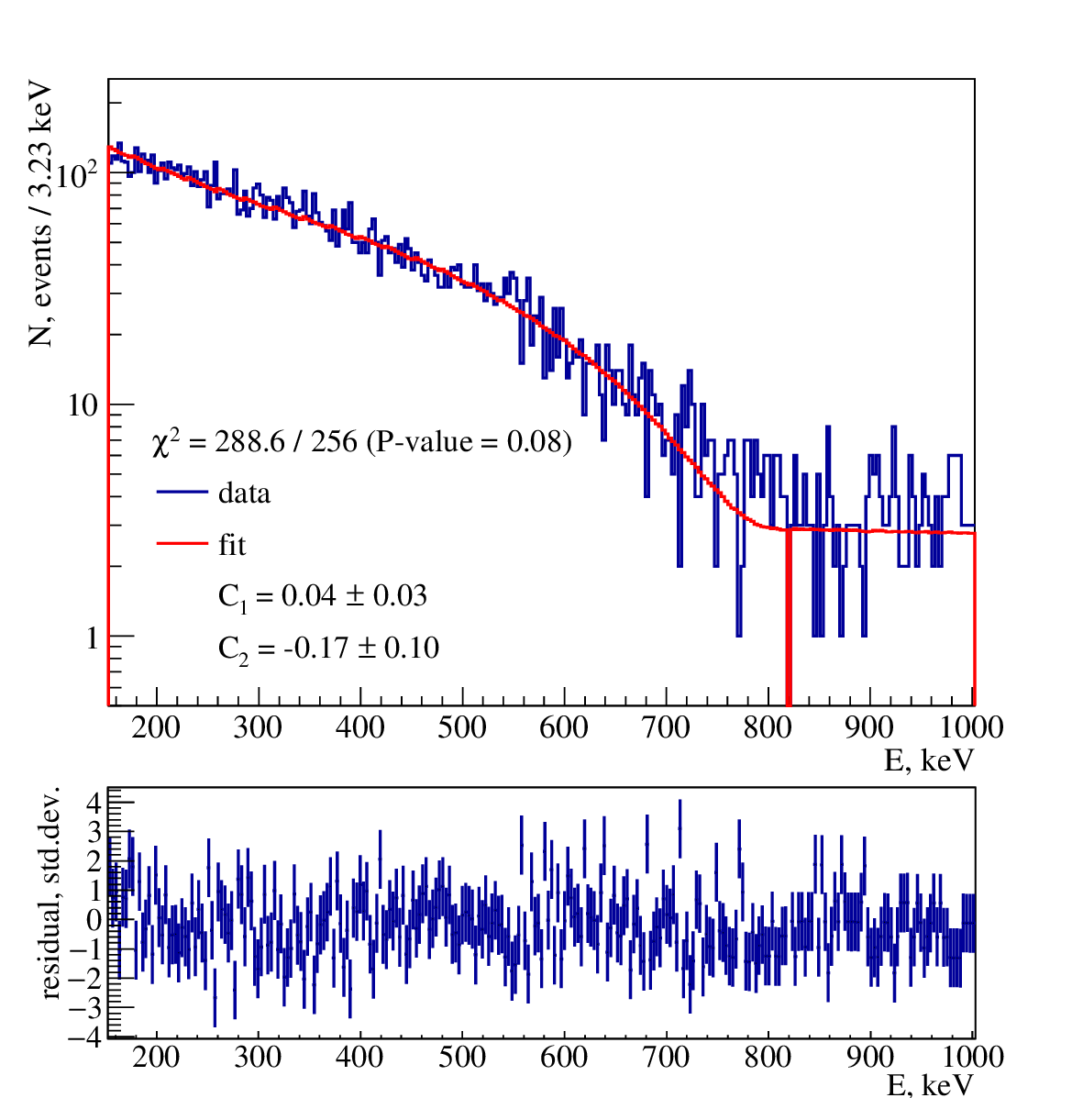}
	\caption{The spectrum of the allowed $\beta-$transition $\rm{{^{144}Pr~(0^-)} \rightarrow {^{144}Nd~(1^-)}}$ measured with the ``target-detector'' spectrometer and the fitting results.}
	\label{Fig4}
\end{figure} 
The measured spectrum is a convolution of the beta spectrum (\ref{eq:beta_basic}) and the function $R(E,W)$ that describes the detector response to an electron with total energy $W$, while $E$ is the energy recorded by the $\beta$-spectrometer.
The response of the detector $R(E,W)$ was modeled by the Monte Carlo method using the GEANT4.10.6 package and G4Em\_Standard\_Physics\_option4 package for the electromagnetic interactions, specially designed for detailed accounting of these interactions at low energies.
\par
The simulation was carried out with a detailed description of the geometry of the installation and additional adjustment of the total thicknesses of the insensitive layers of the detector as well as the distribution of the thickness of the target, reproduced in the shape of the $\alpha$-spectra of $\rm{{^{244}Cm}}$ and $\rm{{^{241}Am}}$ whose small impurities were present in the $\rm{^{144}Ce}$ source. 
As an additional degree of freedom for back-scattered electrons, a variable ratio was introduced for the areas of the response function for electron energy $E$ below and above the original electron energy reduced by 25 keV ($T-25$~keV).
\par
The analysis of the ``target - detector'' experimental data was rather sophisticated as the duration of $^{144}$Ce source storage time by the moment of the experiment begin was already $410$~days, resulting thus in low activity and a complex composition of the background components in the measured spectrum. 
In order to solve this problem, the results of the long-term measurements were divided into two consecutive datasets. 
The analysis of the difference between the first and second parts of the dataset made it possible to significantly reduce the contribution of long-lived background components at the cost of losing statistics from the decays of $^{144}$Ce and~$^{144}$Pr. 
As a result of the measurements, $2024$~one-hour series were selected, which were divided into two equal parts. 
The spectrum corresponding to the first part was fitted by the sum of the second part spectrum, the beta spectra of $^{144}$Ce and $^{144}$Pr and the complementary background components (Fig.~\ref{Fig3}).
\par
The main background component present in the measured spectrum was associated with the activity of the $^{154}$Eu ($\rm{\tau = 12.4~y}$). 
The expected spectrum  of $^{154}$Eu was simulated by the Monte Carlo method similarly to the response function.
The spectrum measured over 1012 hours was fitted with the maximum likelihood method using the $\chi^2$ function (Fig.~\ref{Fig3}). 
The fitting was carried out in the range $(250 - 3100)$~keV, which includes part of the $\beta$-spectrum of $\rm{^{144}Ce}$. 
This made it possible to use the well-established $Q_\beta$-value ($318.7(8)$~keV) of $^{144}$Ce-decay.
\par
The nuclear structure shape factor $C(Z,W)$ is the function searched in the problem of the $\beta$-spectrum measurement. 
The shape factor was parameterized in powers of $W$ similarly to \cite{Lau1956}:
  \begin{equation}
      C(W) = 1 + C_1W + C_2W^{-1}, \label{eq2}
  \end{equation}
 where $C_1$ and $C_2$ are free parameters.
As a result of the analysis, the following expression for the shape factor function was obtained:
\begin{equation}
 C(W) =1 + (-0.023 \pm 0.005)W + (-0.17 \pm 0.09)W^{-1}.\label{eq3}
\end{equation}
The accuracy of the obtained result~(\ref{eq3}) is limited by the weak activity of the $^{144}$Ce source, which requires a small distance from the detector to the target, that leads to large angles of incidence of electrons onto the crystal surface and an increased proportion of backscattered electrons. 
The accuracy in determination of the the nuclear shape factor parameters $C_1$ and $C_2$ could be significantly increased by usage of a more active source with ``target-detector'' spectrometer.

The spectrum of the allowed transition $\rm{{^{144}Pr}~(0^-) \rightarrow {^{144}Nd}~(1^-)}$, fitted similarly with the background component of random coincidences, has shown a good agreement of the model with the experiment (Fig.~\ref{Fig4}).
The fit with fixed $C1=0$ and $C2=0$ gives the minimum $\chi^2$-value is 288.6 for 256 degrees of freedom, which corresponds to the acceptable  $P$-value = 0.08. 
If the parameters $C1$ and $C2$ are free during the fit, their best-fit values are compatible with zero within their uncertainties.

\begin{figure}[t]
	\includegraphics[width=9cm,height=10.5cm]{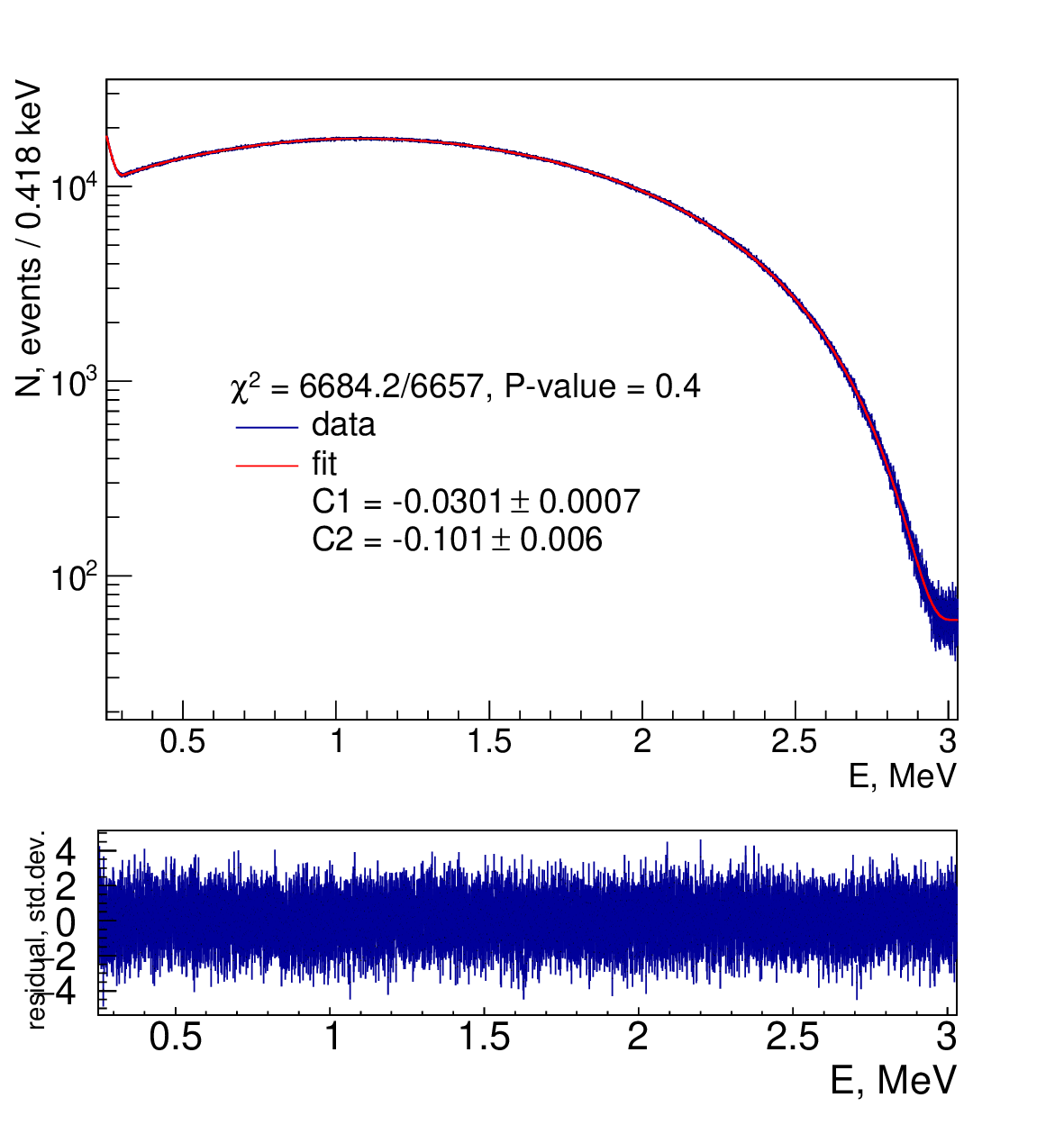}
	\caption{The energy spectrum of the $^{144}$Ce-$^{144}$Pr source measured with the $4\pi-$spectrometer (blue points) and the fitting results (red line). The difference is shown in the units of standard deviations (SD).}
	\label{Fig5}
\end{figure} 
\begin{figure}[t]
	\includegraphics[width=9cm,height=10.5cm]{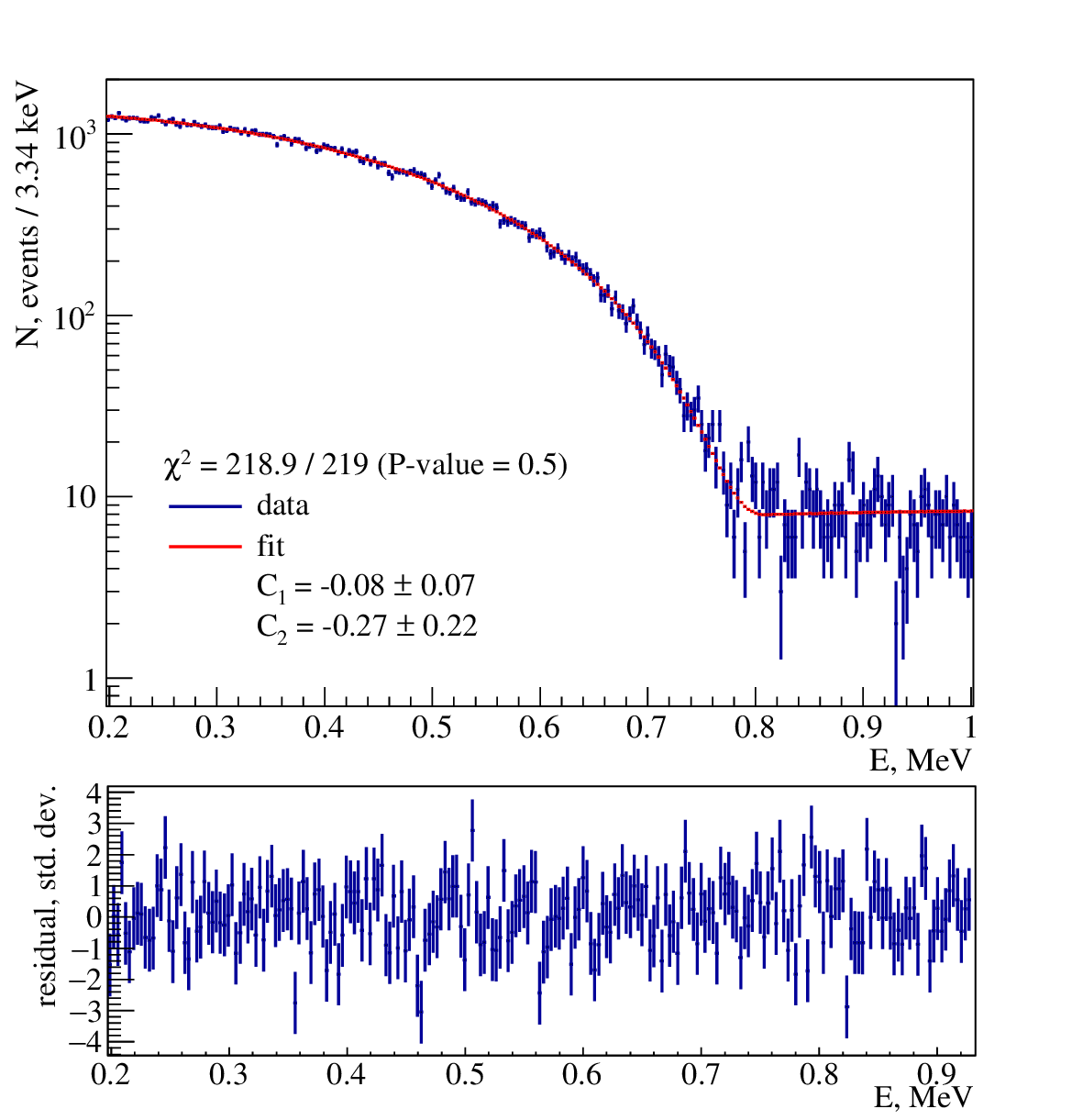}
	\caption{The energy spectrum of the allowed $\beta-$transition ${^{144}\rm{Pr}~(0^-) \rightarrow ^{144}\rm{Nd}~(1^-)}$ measured with the $4\pi-$spectrometer and the fitting results.}
	\label{Fig6}
\end{figure}

\subsection{Analysis of the data of 4$\pi$-spectrometer}
Neglecting the energy of the recoil nucleus, the energy of the electron and neutrino are related by a simple relation $E_e+E_\nu=Q_\beta$.
The $4\pi$ detector geometry makes it possible to measure the real spectrum of electrons and, ideally in the case of delta response $R(E,W)$, allows one to directly determine the antineutrino spectrum from the measured electron one without using the fitting procedure based on~eq.(\ref{eq:beta_basic}). 
However, at electron energies where transitions to excited states of ${\rm{^{144}Pr}}$ or decays of ${\rm{^{144}Ce}}$ contribute to the measured spectrum, corrections must be made for the corresponding branching factors.
\par
Although the response of the $4\pi$ spectrometer is close to the Gaussian function, one should take into account the passage of the insensitive layers of the source and detectors. 
In order to account for these effects detailed Monte Carlo calculations were carried out for different electron energies.

The detector response function $R(E,W)$ determined as a result of fitting looks like as a convolution of a Gaussian function with an exponential tail, which describes small energy losses in the target and non-sensitive layers of the detectors.
As an example, for 1 MeV electrons the standard deviation for response function (the square root of the second central moment $\sigma^2$) is 56.8 keV and the fraction of electrons detected with an energy less than 0.850 MeV is 3.4\%.

The response function calculated by the Monte Carlo method was described analytically as:
\begin{equation}
     R(E,W) = \exp(P_0(W) + E/P_1(W)) \theta(T-E).
    \label{eq4}
\end{equation} 
The function $P_0(W)$ was estimated from the condition of conservation of the full integral of the response function. $P_1(W)$ describes the dependence of the standard deviation of the response function $\sigma_{RF}(W)$ with two additional free parameters $p_1$ and $p_2$:
$\sigma_{RF}(W) = \sigma_{MC}(W)(1+p_1W+p_2W^2)$, where the dependence $\sigma_{MC}(W)$ is obtained by the Monte Carlo method.
The parameters $p_1$, $p_2$ are responsible for imperfections in the modeling, $\theta(x)$ is the Heaviside step function.

The spectrum of $\rm{^{144}C-^{144}Pr}$ source measured by $4\pi$ $\beta$-spectrometer over 106 hours is shown in Fig.~\ref{Fig5}.
The fitting  was carried out by the maximum likelihood method with the $\chi^2$  likelihood function with the shape factor $C(Z,W)$ similar to the one (\ref{eq2}) used in the data analysis of the ``target-detector'' spectrometer. 
\par
As a result of the analysis, the following values of the parameters $C_1$ and $C_2$ of the nuclear shape factor function were obtained:
  \begin{equation}
    C(W) = 1 + (-0.0301 \pm 0.0007)W + (-0.101 \pm 0.006)W^{-1}.
  \label{eq5}
  \end{equation}

From (\ref{eq3}) and (\ref{eq5}) one could see that the results for the parameters $C_1$ and $C_2$ obtained with two different beta spectrometers agree with each other within the uncertainties (Fig.\ref{Fig3} and Fig.\ref{Fig5}).
The achieved accuracy of $C_1$ and $C_2$ determination in the measurement with the $4\pi$ spectrometer is significantly higher than for the "target detector" setup. 
The fitting of the allowed transition measured spectrum has shown a good agreement of the theoretical model used with the experiment as in the case of the measurement with the target-detector spectrometer (Fig.~\ref{Fig6}). 
\par
The $\rm{^{144}Pr}$ $\beta$-spectrum has been previously measured in several works \cite{Lau1956, Gra1958, Por1959, Dan1966, Nag1971} that used different expressions for the $F(Z,W)$ Fermi function  and for the $C(Z,W)$ shape factor.
Therefore, a direct comparison of the results obtained is difficult.
Nevertheless, our result for $C(Z,W)$ is more precise but consistent  with \cite{Nag1971,Dan1966} within experimental uncertainties.

\section{NEUTRINO SPECTRUM}
Since the $\beta$-spectrum shape measured by the $4\pi$-spectrometer turns out to be more accurate, it was used to determine the neutrino spectrum.
Provided there is a measured shape of each component of $\beta$-spectrum one could evaluate a neutrino spectrum though conservation of energy in the decay as the total energy of each decay is given by its endpoint energy. 
The neutrino spectrum $N(E_\nu)$ is obtained from (\ref{eq:beta_basic}) by the replacements $W$ with $E_\nu=(W_0-W)$.
\par
The main interest of sterile neutrino experiments lies above the energy threshold of inverse beta decay on hydrogen, both in terms of shape and spectrum fraction. 
As for ${\rm{^{144}Ce}}$--${\rm{^{144}Pr}}$ source, only two $\beta$-transitions contribute to the neutrino spectrum above this threshold: ${^{144}\rm{Pr}}(0^-) \rightarrow {^{144}\rm{Nd}}(0^+)$ and $\rm{{^{144}Pr}(0^-) \rightarrow {^{144}Nd}(2^+)}$ with the branching ratios of $97.9$\% and $1.04$\%, respectively. 
\par
\begin{figure}[t]
\center{\includegraphics[width = 9cm,height = 10.5cm]{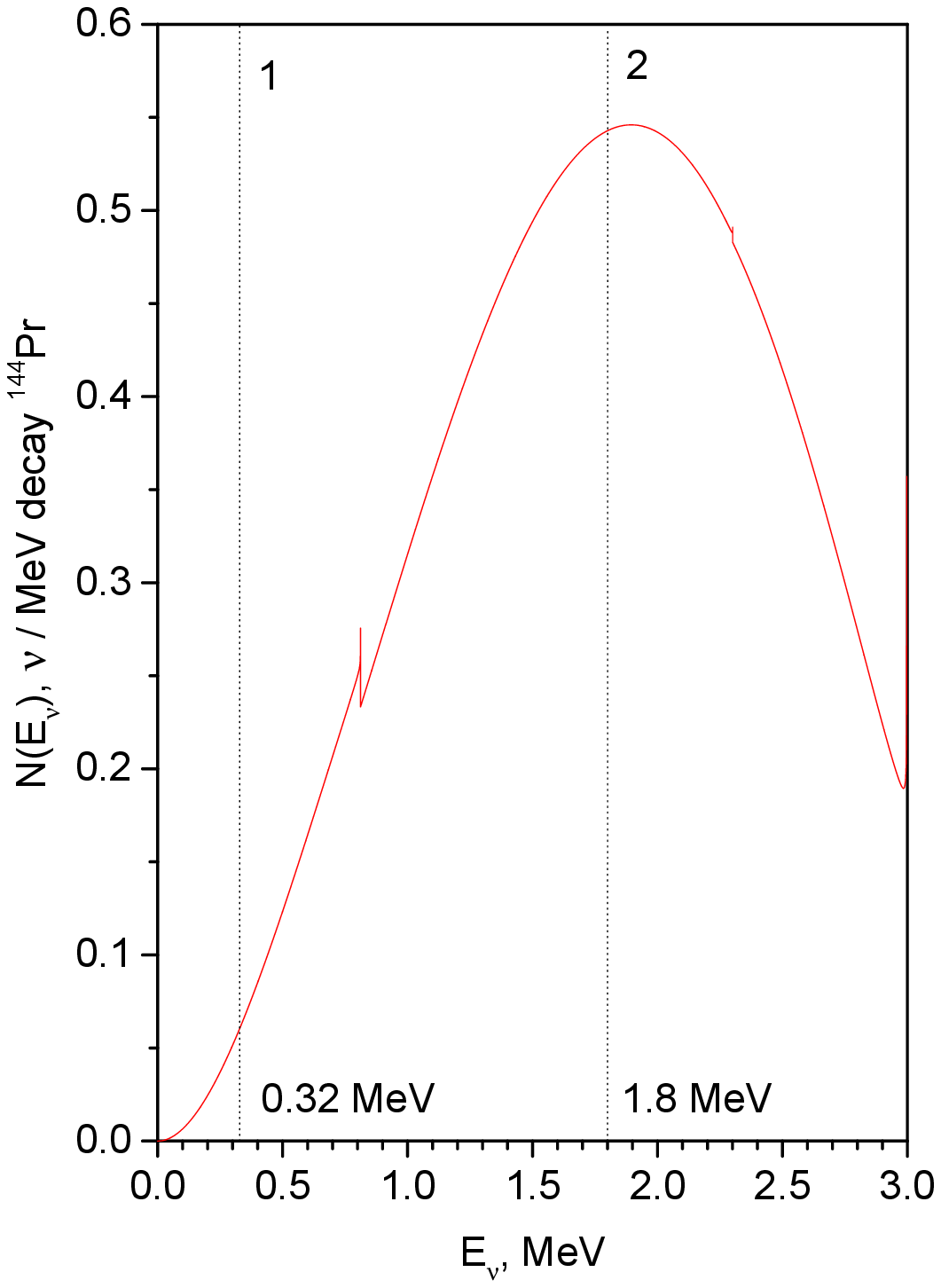}}
\caption{The energy spectrum of electron antineutrinos of the $^{144}$Pr source. Numerical values for $E_\nu \geq 1.8$~MeV are given in Tab. \ref{tab1}.} 
\label{Fig7}
\end{figure}

Fig.~\ref{Fig7} illustrates the spectrum $N(E_\nu)$ of electron antineutrinos from $^{144}$Pr source. 
The numerical values $N(E_\nu)$ for neutrinos with energies above the IBD threshold are shown in the Tab.~\ref{tab1}.
The uncertainties of $N(E_\nu)$ are calculated with toy Monte Carlo method based on the  uncertainties of the evaluated nuclear shape factor parameters $C_1$ and $C_2$ from (\ref{eq5}) and taking into account that these parameters are in quite strong anticorrelation with correlation coefficient of 0.973. The parameters evaluated with this method were simulated 10000 times and used for performing 2-dimensional spectrum density plot. The RMS of density distributions of its monoenergetic projections was taken as the statistical uncertainty of the spectrum.

The uncertainty of the energy scale $E_\nu$ is determined by the accuracy of the $Q_\beta$-value of $\beta$-transition of the $\rm{^{144}Pr}$ to the ground state of $\rm{^{144}Nd}$ \cite{iaea}.

Neutrinos from the decay of $\rm{^{144}Ce}$ have the same intensity as $\rm{^{144}Pr}$-neutrinos, but their energies are less than 318 keV (the region to the left of line 1 in Fig.~\ref{Fig7}).
The threshold value of the IBD reaction on hydrogen in the laboratory system is $E_{thr}\rm{=1.806~MeV}$ (line 2). 
\par
\setlength\extrarowheight{3.5pt}

\begin{table}[ht]
\centering
\scriptsize
\caption{Neutrino spectrum data ($N(E_\nu)$) for energies above the threshold of IBD reaction. The error values ($1\sigma$) are shown in parentheses.}
\begin{tabular}{>{\centering\arraybackslash}p{0.9cm}>{\centering\arraybackslash}p{1.5cm}>{\centering\arraybackslash}p{0.9cm}>{\centering\arraybackslash}p{1.5cm}>{\centering\arraybackslash}p{0.9cm}>{\centering\arraybackslash}p{1.5cm}}
\hline
E, keV & N, 1/MeV & E, keV & N, 1/MeV &  E, keV & N, 1/MeV  \\
\hline
1809 & 0.54431(80) & 2209 & 0.51117(79) & 2609 & 0.36696(74) \\
1819 & 0.54488(80) & 2219 & 0.50887(79) & 2619 & 0.36235(73) \\
1829 & 0.54538(80) & 2229 & 0.50649(79) & 2629 & 0.35770(73) \\
1839 & 0.54581(80) & 2239 & 0.50405(79) & 2639 & 0.35301(73) \\
1849 & 0.54618(80) & 2249 & 0.50154(79) & 2649 & 0.34828(73) \\
1859 & 0.54647(80) & 2259 & 0.49896(79) & 2659 & 0.34351(72) \\
1869 & 0.54669(80) & 2269 & 0.49632(78) & 2669 & 0.33870(72) \\
1879 & 0.54684(80) & 2279 & 0.49363(78) & 2679 & 0.33386(72) \\
1889 & 0.54692(80) & 2289 & 0.49092(78) & 2689 & 0.32899(72) \\
1899 & 0.54692(80) & 2299 & 0.48857(78) & 2699 & 0.32408(71) \\
1909 & 0.54686(80) & 2309 & 0.48096(78) & 2709 & 0.31914(71) \\
1919 & 0.54672(80) & 2319 & 0.47803(78) & 2719 & 0.31418(71) \\
1929 & 0.54652(80) & 2329 & 0.47504(78) & 2729 & 0.30919(71) \\
1939 & 0.54624(80) & 2339 & 0.47198(78) & 2739 & 0.30418(70) \\
1949 & 0.54588(80) & 2349 & 0.46886(78) & 2749 & 0.29914(70) \\
1959 & 0.54546(80) & 2359 & 0.46567(78) & 2759 & 0.29408(70) \\
1969 & 0.54496(80) & 2369 & 0.46242(77) & 2769 & 0.28901(69) \\
1979 & 0.54439(80) & 2379 & 0.45910(77) & 2779 & 0.28392(69) \\
1989 & 0.54375(80) & 2389 & 0.45573(77) & 2789 & 0.27881(69) \\
1999 & 0.54303(80) & 2399 & 0.45229(77) & 2799 & 0.27370(68) \\
2009 & 0.54224(80) & 2409 & 0.44878(77) & 2809 & 0.26857(68) \\
2019 & 0.54138(80) & 2419 & 0.44522(77) & 2819 & 0.26345(68) \\
2029 & 0.54045(80) & 2429 & 0.44160(77) & 2829 & 0.25832(67) \\
2039 & 0.53944(80) & 2439 & 0.43792(76) & 2839 & 0.25320(67) \\
2049 & 0.53836(80) & 2449 & 0.43417(76) & 2849 & 0.24808(67) \\
2059 & 0.53720(80) & 2459 & 0.43037(76) & 2859 & 0.24298(66) \\
2069 & 0.53598(80) & 2469 & 0.42652(76) & 2869 & 0.23790(66) \\
2079 & 0.53467(80) & 2479 & 0.42260(76) & 2879 & 0.23284(65) \\
2089 & 0.53330(80) & 2489 & 0.41863(76) & 2889 & 0.22781(65) \\
2099 & 0.53185(80) & 2499 & 0.41461(76) & 2899 & 0.22283(65) \\
2109 & 0.53033(80) & 2509 & 0.41053(75) & 2909 & 0.21789(64) \\
2119 & 0.52874(79) & 2519 & 0.40640(75) & 2919 & 0.21303(64) \\
2129 & 0.52708(79) & 2529 & 0.40221(75) & 2929 & 0.20826(64) \\
2139 & 0.52534(79) & 2539 & 0.39798(75) & 2939 & 0.20360(63) \\
2149 & 0.52353(79) & 2549 & 0.39369(75) & 2949 & 0.19910(63) \\
2159 & 0.52165(79) & 2559 & 0.38935(75) & 2959 & 0.19483(62) \\
2169 & 0.51970(79) & 2569 & 0.38497(74) & 2969 & 0.19096(62) \\
2179 & 0.51767(79) & 2579 & 0.38053(74) & 2979 & 0.18799(62) \\
2189 & 0.51558(79) & 2589 & 0.37606(74) & 2989 & 0.18833(62) \\
2199 & 0.51341(79) & 2599 & 0.37153(74) & 2995 & 0.20092(80) \\

\hline
\end{tabular}
\label{tab1}
\end{table}
\normalsize
Considering the precision given by our measurement one could apply toy Monte Carlo to convert the uncertainties on the shape factor parameters to the fraction of the $\rm{^{144}Pr}$ $\beta$-spectrum above the threshold IBD reaction that is 
$(0.50057 \pm 0.00006_{stat}\pm 0.00069_{syst})$,
so the statistic precision of this value is as low as 0.012~\%.
There are two main causes of systematic errors. 
The first uncertainty evaluated as the difference of the values derived for two boundary cases of low-energy screening correction is 0.00022 (0.044~\%).
A more significant systematic error is associated with the 1.2~keV uncertainty of the energy scale at 1.806 MeV which results in a value of 0.00065 (0.13~\%).
\par
\begin{figure}[t]
\center{\includegraphics[width = 9cm,height = 10.5cm] {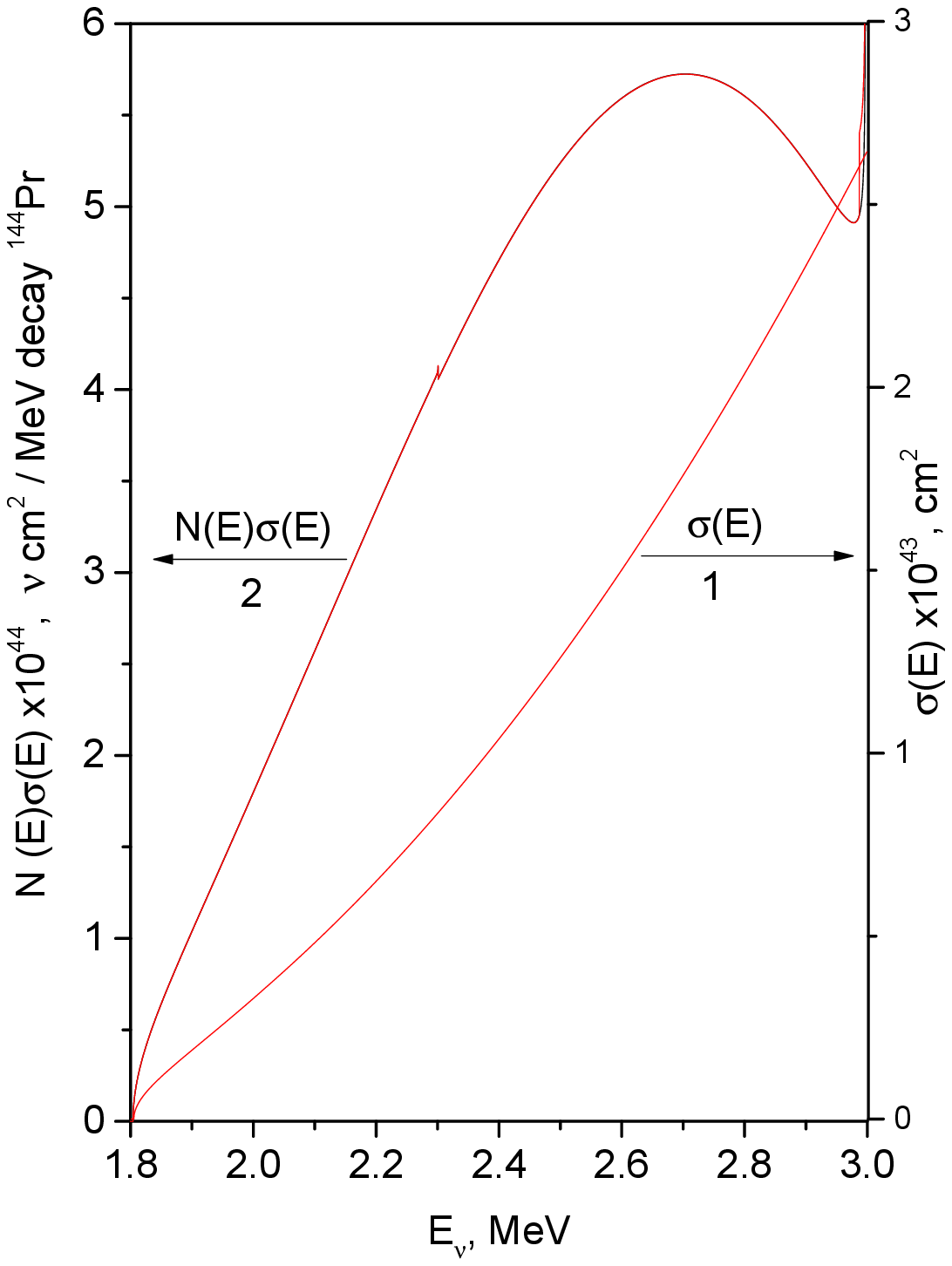}}
\caption{The crossection of inverse beta-decay ($\sigma(E_\nu)$) and a product of $\bar{\nu}$-spectra of the $^{144}$Pr ($N(E_\nu)$) and the crossection.} \label{Fig8}
\end{figure}

The total cross section of the IBD reaction is determined by the integral from the threshold $E_{thr}$ to the maximum neutrino energy $E_0$ over the product of the IBD reaction cross section $\sigma(E_\nu)$ and the neutrino flux $N(E_\nu)$ (Fig.\ref{Fig8}):
\begin{equation}
\sigma_{144Pr} = \int_{E_{thr}}^{E_0} \sigma(E_\nu)\,N(E_\nu)\,dE_\nu.
\label{CStotal}
\end{equation}
Using the cross section of IBD reaction from ~\cite{Str2003,Ric2022,Ric2025,Add2025}, one could evaluate the integrated cross section, that is 
$\sigma_{144Pr}=(4.7344 \pm 0.0006_{stat} \pm 0.013_{syst}) \times 10^{-44}$~cm$^2$ per one decay of $\rm{^{144}Pr}$ nucleus.
The total systematic error reported here is obtained by taking into account 
the uncertainty in the screening correction ($\sigma$=0.0041 (0.09\%)) and  inaccuracy of the energy scale $E_\nu$ ($\sigma$=0.012 (0.25\%)).
The latter can be substantially improved by any refinement of the $Q_\beta$-value of $\rm{^{144}Pr}$-decay.
The obtained precision of $\sigma_{144Pr}$ in $0.25$\% would correspond to an experimental uncertainty of neutrino count rate expected for a sterile  neutrino experiment using $\rm{^{144}Ce - ^{144}}$Pr source. 
In this case this value would be sub-dominated by other experimental uncertainties, e.g. the precision of the source calorimetry for Borexino-SOX experiment has reached precision of $0.4$\%~\cite{Alt2016}.
\par

We can also provide an analytical interpolation of the spectrum, that could be expressed in two energy intervals. 
For neutrino energies $E$ above the  unique forbidden transition endpoint of $2.301$~MeV, the spectrum is interpolated as follows ($E$ in MeV units):
\begin{eqnarray}
\nonumber N(E)_{(2.3)\rm{MeV}}=a_1+a_2 E + a_3 E^2 + a_4 E^3\\ 
\nonumber+ \exp(a_5+a_6 (2.9974 - E))+\exp(a_7+a_8 (2.9974-E)).\\
\end{eqnarray}

Between the inverse beta decay threshold of $1.806$~MeV and $2.301$~MeV, we also include a transition to the excited level of $0^-\rightarrow 2^+ $ in $^{144}$Nd, adding an additional component $N(E)_{(2.3)\text{MeV}}$:
\begin{eqnarray}
\nonumber N(E)_{(1.8)\text{MeV}}=N(E)_{(2.3)\text{MeV}}+a_{9}+a_{10} E + a_{11} E^2\\ 
\nonumber + a_{12} E^3 + \exp(a_{13}+a_{5}+a_{6} (2.3010 - E))+ \\ \nonumber+\exp(a_{13}+a_{7}+a_{8} (2.3010 - E)),\\
\end{eqnarray}
where the parameters are defined according to the Tab.~\ref{tab2}. 
The accuracy of this analytical interpolation could be estimated by the ratio of two integrals: the product of the measured spectrum with the cross section of the inverse beta decay and the product of the interpolation function with the same cross section. 
For neutrino energies greater than $1.8$~MeV, the deviation of these two values was found to be as low as $0.06$~\%.
\par

\setlength\extrarowheight{3.5pt}
\begin{table}[ht]
	\centering
	\footnotesize
	\caption{Parameters of neutrino spectrum analytical description above $1.8$~MeV.}
	\begin{tabular}{|>{\centering\arraybackslash}p{0.8cm}|>{\centering\arraybackslash}p{2.5cm}|>{\centering\arraybackslash}p{0.8cm}|>{\centering\arraybackslash}p{2.5cm}|}
		\firsthline
		$a_1$ & $-5.1796$  & $a_{8}$   & $-2536.09$  \\
        $a_2$ & $ 6.7996$  & $a_{9}$   & $4.38127$  \\
        $a_3$ & $ -2.5903$ & $a_{10}$  & $-5.36022$ \\
        $a_4$ & $ 0.30622$ & $a_{11}$  & $2.19506$  \\
        $a_5$ & $ -3.4667$ & $a_{12}$  & $-0.300772 $\\
        $a_6$ & $ -131.65$  & $a_{13}$  & $-3.6471$ \\
        $a_7$ & $ -2.0722$ & &  
        

\\
		\hline
	\end{tabular} 
	\label{tab2}
\end{table}
\normalsize

\section{CONCLUSION}
The $\beta$-spectra of the $\rm{{^{144}Ce}-{^{144}Pr}}$ source were studied using two spectrometers based on semiconductor Si(Li) detectors. 
The spectra were measured using a beta spectrometer in the classical ``target-detector'' scheme and with an original $4\pi$ full absorption spectrometer consisting of two of Si(Li) detectors. 
The models of the beta spectrometer response function calculated through the Monte Carlo simulation were tested by fitting the spectrum of the allowed beta transition $^{144}\rm Pr~(0^-) \rightarrow {^{144}\rm Nd}~(1^-)$ and have shown good agreement between the response models and experimental data.
The function of the nuclear shape factor of the ground state beta transition in ${^{144}\rm{Pr}}-{^{144}\rm{Nd}}$ has been defined as 
$C(W) = 1 + (-0.0301 \pm 0.0007)W + (-0.101 \pm 0.006)W^{-1}$.
The electron antineutrino spectrum of the $\rm{{^{144}Pr}}$ decays was obtained from the measured $\beta$-spectrum and the $C(W)$ shape factor parameters. 
The reduced cross section for the inverse beta decay reaction for the $\rm{{^{144}Pr}}$ electron anti-neutrino source is $(4.7344 \pm 0.0006_{stat} \pm 0.013_{syst}) \times 10^{-44}$~cm$^2$ per ${\rm{^{144}Pr}~decay}$ which is the most accurate measurement up to date.

\section*{ACKNOWLEDGEMENTS}
This work has been supported by Russian Science Foundation (project no.24-12-00046).


\end{document}